# Parametric Amplification of a Terahertz Quantum Plasma Wave

S. Rajasekaran[1], E. Casandruc[1], Y. Laplace[1], D. Nicoletti[1], G. D. Gu[4], S. R. Clark[3,1], D. Jaksch[2,5], A. Cavalleri[1,2]

[1]Max Planck Institute for the Structure and Dynamics of Matter, Luruper Chaussee 149, 22761 Hamburg, Germany
[2]Department of Physics, Oxford University, Clarendon Laboratory, Parks Road, OX1 3PU Oxford, United Kingdom
[3]Department of Physics, University of Bath, Claverton Down, BA2 7AY, Bath United Kingdom
[4]Condensed Matter Physics and Materials Science Department, Brookhaven National Laboratory, Upton, New York, USA
[5]Centre for Quantum Technologies, National University of Singapore, 3 Science Drive 2, Singapore 117543, Singapore

**Many applications in photonics require all-optical manipulation of plasma waves[1], which can concentrate electromagnetic energy on sub-wavelength length scales. This is difficult in metallic plasmas because of their small optical nonlinearities. Some layered superconductors support weakly damped plasma waves[2,3], involving oscillatory tunneling of the superfluid between capacitively coupled planes. Such Josephson plasma waves (JPWs) are also highly nonlinear[4], and exhibit striking phenomena like cooperative emission of coherent terahertz radiation[5,6], superconductor-metal oscillations[7] and soliton formation[8]. We show here that terahertz JPWs in cuprate superconductors can be parametrically amplified through the cubic tunneling nonlinearity. Parametric amplification is sensitive to the relative phase between pump and seed waves and may be optimized to achieve squeezing of the order parameter phase fluctuations[9] or single terahertz-photon devices.**

Cuprates are strongly anisotropic superconductors in which transport is made three-dimensional by Josephson tunneling between the Cu-O planes. Tunneling reduces the superfluid density in the direction perpendicular to the planes and hence the frequency of the plasmon to below the average pair breaking gap. Weakly damped oscillations of the superfluid sustain transverse Josephson Plasma Waves (JPWs) that propagate along the planes.

Consider a complex superconducting order parameter in the $i^{th}$ Cu-O plane $\psi_i(x,y,t) = |\psi_i(x,y,t)| e^{i\theta_i(x,y,t)}$, which depends on two in-plane spatial coordinates $x$ and $y$ and on time $t$. For a THz-frequency optical field polarized perpendicular to the planes, excitations above the superconducting gap are negligible and the modulus of the order parameter $|\psi|^2$ (number of Cooper pairs) is nearly constant in space and time. Hence, the electrodynamics is dominated by the order-parameter phase $\theta_i(x,y,t)$. Ignoring at first the spatial dependence of the phase, the local tunneling strength can, from the Josephson equations[10], be expressed in terms of an equivalent inductance, $L$, which depends on the local interlayer phase difference $\theta_{i,i+1}(t) = \theta_i(t) - \theta_{i+1}(t)$ as $L(\theta_{i,i+1}(t)) \sim L_0 / \cos(\theta_{i,i+1}(t))$ ($i$ and $i+1$ are the indices for two neighboring layers). Here $L_0 = \dfrac{\hbar}{2eI_c}$ is the inductance at equilibrium, $\hbar$ the reduced Planck's constant, $2e$ the Cooper pair charge and $I_c$ the critical current. Denoting the capacitance of the Cu-O planes with a constant $C$, we express the Josephson Plasma Resonance (JPR) frequency as

$$\omega_{JP}{}^2 = \frac{1}{L(\theta_{i,i+1}(t))C} = \omega_{JP0}{}^2 \cos[\theta_{i,i+1}(t)], \text{ where } \omega_{JP0}{}^2 = \frac{1}{L_0 C}$$ is the equilibrium value.

Correspondingly, the oscillator strength $f$ for the plasma oscillations[11] is also a function of the interlayer phase and scales as $f = f_0 \cos[\theta_{i,i+1}(t)]$. The dependence of the oscillator strength $f$ on the cosine of the superconducting phase corresponds to a third order optical nonlinearity.

According to the second Josephson equation[10], the interlayer phase difference $\theta_{i,i+1}(t)$ advances in time with the time integral of the interlayer voltage drop, as $\frac{\partial[\theta_{i,i+1}(t)]}{\partial t} = \frac{2eV}{\hbar}$. For an optical field made resonant with the Josephson plasma frequency $E(t) = E_0 \sin(\omega_{JP0} t)$, the interlayer phase oscillates as $\theta_{i,i+1}(t) = \theta_0 \cos(\omega_{JP0} t)$, where $E_0$ is the field amplitude and $\theta_0 = \frac{2ed}{\hbar \omega_{JP0}} E_0$ ($d \sim 1$ nm is the interlayer distance). This implies that the oscillator strength

$$f_{i,i+1}(t) = f_0 \cos(\theta_0 \cos(\omega_{JP0} t)) \approx f_0 \left(1 - \frac{\theta_0^2 + \theta_0^2 \cos(2\omega_{JP0} t)}{4}\right)$$ is modulated at a frequency $2\omega_{JP0}$, whenever the field $E_0$ is large enough to make the phase excursion $\theta_0$ sizeable.

Figure 1 provides a pictorial representation of this physics. We plot a vector that represents both the *phase difference* $\theta_{i,i+1}(t)$ (vector angle) and the *oscillator strength* $f_{i,i+1}(t)$ (vector length). This picture shows how, for small driving fields, only $\theta_{i,i+1}(t)$ oscillates at the driving frequency $\omega_{JP0}$, whereas for larger fields these oscillations are accompanied by a $2\omega_{JP0}$ modulation of the oscillator strength $f_{i,i+1}(t)$.

Note also that the phenomena discussed above can be casted in terms of a Mathieu equation (see Supplementary Section S1). Thus, a $2\omega_{JP0}$ modulation of the oscillator strength can serve as a pump for the parametric amplification of a second, weak plasma wave at frequency $\omega_{JP0}$. In this paper we demonstrate experimentally this effect in La$_{1.905}$Ba$_{0.095}$CuO$_4$ (LBCO$_{9.5}$), a cuprate superconductor with the equilibrium JPR at $\omega_{JP0} \cong 0.5$ THz.

Terahertz pulses, generated with a photoconductive antenna[12], were used as a weak probe of JPWs. A typical THz-field trace[13] reflected by the sample is shown in Fig. 2A. Two different measurements are displayed: one taken below (red line) and the other one above (black line) the superconducting transition temperature $T_c$ = 32 K. In the superconducting state, long-lived oscillations with ~ 2 ps period were observed on the trailing edge of the pulse, indicative of the JPR at $\omega_{JP0} \cong 0.5$ THz. Figure 2B (solid red line) displays the corresponding reflectivity edge in frequency domain. The solid lines in Fig. 2C -2D are the complex dielectric permittivity $\varepsilon(\omega)$ and the loss function $L(\omega) = -\text{Im}\left(\dfrac{1}{\varepsilon(\omega)}\right) = \dfrac{\varepsilon_2(\omega)}{(\varepsilon_1(\omega) + \varepsilon_2(\omega))^2}$. $L(\omega)$ peaks at $\omega_{JP0}$, where the real part of the dielectric permittivity, $\varepsilon_1(\omega)$, crosses zero.

These optical properties could be well reproduced by solving the wave equation in the superconductor in one dimension[8] (see Supplementary Section S2), which yields the space and time dependent order parameter phase $\theta_{i,i+1}(x,t)$ (Fig. 2E) and the corresponding changes (negligible in linear response regime) of the oscillator strength $f = f_0 \cos[\theta_{i,i+1}(x,t)]$ (Fig. 2F). The reflectivity, complex permittivity, and loss function (dashed lines in Fig. 2B, 2C, and 2D, respectively), calculated from these simulations by solving the electromagnetic field at the sample surface[8], are in good agreement with the experimental data.

Amplification of a weak JPW like the one above (*probe* field) was achieved by mixing it with a second, intense *pump* field, which resonantly drove the Josephson phase to large amplitudes. Quasi-single cycle THz pulses, generated in LiNbO$_3$ with the tilted pulse front method[14] (yielding field strengths up to ~100 kV/cm), were used to excite these waves in nonlinear regime. The spectral content of these pulses extended between 0.2 and 0.7 THz, centered at the JPR frequency (see Supplementary Section S3). Note that the pump field strength used in this experiment exceeds the expected threshold to access the nonlinear regime, defined by $\theta_0 = \frac{2eE_0 d}{\hbar \omega_{JP0}} \sim 1$ and corresponding in this material to field amplitudes $E_0 = \frac{\hbar \omega_{JP0}}{2ed}$ ~ 20 kV/cm.

In Fig. 3, we report the time-delay dependent, spectrally integrated pump-probe response of LBCO$_{9.5}$. Changes in the reflected probe field were measured at one specific point along the internal delay trace of the probe, as a function of pump-probe time delay. For a system in which the optical properties are dominated by a single plasma resonance, the spectrally integrated response is proportional to the plasma oscillator strength *f*.

As shown in Fig. 3A-3B, this integrated response exhibits a reduction of the reflectivity and oscillations at a frequency $\sim 2\omega_{JP0}$. Note that the oscillation frequency did not depend on the pump electric field strength $E_0$, while the frequency reduced when the base temperature of the experiment was increased, consistent with the reduction of the equilibrium $\omega_{JP0}$ (see Supplementary Sections S4 and S5). The effect completely disappeared at $T > T_c$.

Hence, the theoretically predicted $2\omega_{JP0}$ modulation of the total oscillator strength $f$ (see above) is well reproduced by the data in Fig. 3. This response could also be simulated using the space- and time-dependent sine-Gordon equation (see Fig. 3C-3D). Good agreement between experiment and theory was obtained (see dashed lines in Fig. 3A-3B).

Selected time-domain probe traces measured before and after excitation are displayed in Fig. 4. Crucially, at specific time delays the probe field is amplified (Fig. 4A), whereas at other delays it is suppressed (Fig. 4B) with respect to that measured at equilibrium.

In Fig. 5 we report the time-delay and frequency dependent loss function

$$L(t,\omega) = -\text{Im}\left(\frac{1}{\varepsilon(t,\omega)}\right) = \frac{\varepsilon_2(t,\omega)}{(\varepsilon_1(t,\omega) + \varepsilon_2(t,\omega))^2},$$

a quantity that peaks at the zero crossing of $\varepsilon_1(t,\omega)$ and is always positive for a dissipative medium (i.e., a medium with $\varepsilon_2(t,\omega) > 0$). The experimental data of Fig. 5A show that after excitation $L(\omega)$ acquires negative values around $\omega_{JP0}$ (red regions). This is indicative of a negative $\varepsilon_2(t,\omega)$ and hence amplification. The effect is strong near zero pump-probe time delay, then disappears after ~1 ps and is observed again periodically with a repetition frequency of $\sim 2\omega_{JP0}$. The same effect appears also in the simulations (Fig. 5B), yielding periodic amplification at a repetition frequency of $2\omega_{JP0}$.

The data and theoretical analysis reported here demonstrate that terahertz JPWs can be parametrically amplified, exhibiting the expected sensitivity to the relative phase of strong and weak fields mixed in this process and the oscillatory dependence at twice the frequency of the drive. This effect is of interest for applications in photonics or as a phase-coherent nonlinear probe of the superfluid itself[15]. Moreover, the ability to amplify a plasma wave could lead to single-THz photon manipulation devices that may operate above 1 K temperatures. These would exploit concepts that to date have been developed only at microwave frequencies and in the milli-Kelvin regime[16–19]. Finally, the parametric phenomena discussed here can also potentially be used to squeeze[9,20, 21] the superfluid phase, and may lead to control of fluctuating superconductivity[22], perhaps even over a range of temperatures above $T_c$ [23,24].

# Methods

Laser pulses with 100 fs duration and ~5 mJ energy from a commercial Ti:Sa amplifier were split into 3 parts (92%, 7%, 1%). The most intense beam was used to generate strong-field THz pulses with energies up to ~3 µJ via optical rectification in LiNbO$_3$ with the tilted pulse front technique. These were collimated and then focused at normal incidence onto the sample (with polarization perpendicular to the Cu-O planes, i.e. along the $c$ axis) using a Teflon lens and a parabolic mirror, with focal lengths of 150 mm and 75 mm, respectively. The pump field at the sample position was calibrated with electro-optic sampling in a 0.2-mm-thick GaP crystal, yielding a maximum value of ~100 kV/cm (see also Supplementary Section S3).

The 7% beam was used to generate the THz probe pulses with a photoconductive antenna. These had a dynamic bandwidth of 0.1-3 THz, corresponding to a time resolution of ~250 fs. The $c$-axis optical properties of the superconductor (both at equilibrium and throughout the pump-induced dynamics) were probed in reflection geometry, with a probe incidence angle of 45°. The reflected probe pulses were electro-optically sampled in a 1-mm-thick ZnTe crystal, using the remaining 1% of the 800 nm beam. This measurement procedure returned the quantity E(t, τ), with $t$ being the pump-probe delay and τ being the internal electro-optic sampling time coordinate.

The complex optical properties of the superconductor at equilibrium were determined by measuring the complex-valued E(ω) (pump off) both at T < $T_c$ and T > $T_c$ and by referencing it to the normal-state reflectivity measured in the same batch of samples[25].

In the spectrally integrated pump-probe traces of Fig. 3, E(t,τ) was measured by scanning the pump-probe delay $t$ at a fixed internal delay τ. This was chosen to be on the trailing edge of the pulse, where the JPR oscillations are present. Note that the observed dynamics, and in particular the 2ω$_{JP0}$ oscillations, did not depend significantly on the specific τ value at which the scan was performed.

The frequency and time-delay dependent loss function of Fig. 5 (as well as all complex optical properties of the perturbed material) was determined by applying Fresnel equations[11] to the pump-induced changes in the reflected electric field. These were normalized by independently recording E(t,τ) in presence and absence of the THz pump field. Note that there was no need to take into account any pump-probe penetration depth mismatch in the calculation.

In the simulations, the Josephson phase evolution $\theta_{i,i+1}(x,t)$ was determined through the one-dimensional sine-Gordon equation[6]:

$$\frac{\partial^2 \theta_{i,i+1}(x,t)}{\partial x^2} - \frac{1}{\gamma}\frac{\partial \theta_{i,i+1}(x,t)}{\partial t} - \frac{\varepsilon_r}{c^2}\frac{\partial^2 \theta_{i,i+1}(x,t)}{\partial t^2} = \frac{\omega_{JP0}^2 \varepsilon_r}{c^2}\sin\theta_{i,i+1}(x,t)$$

being $\gamma$ a damping constant, $c$ the speed of light, $\varepsilon_r$ the dielectric permittivity, and $\omega_{JP0}$ the equilibrium JPR frequency. This equation was solved numerically, with the THz pump and probe fields overlapping at the vacuum-superconductor interface. For more details on this topic, we refer the reader to Supplementary Section S2.

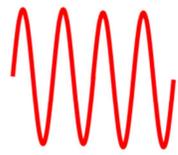
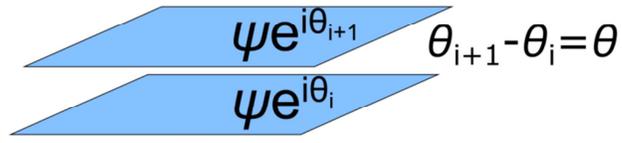
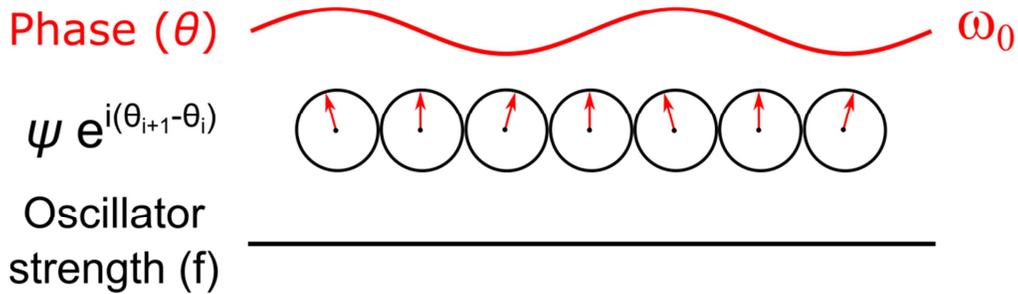
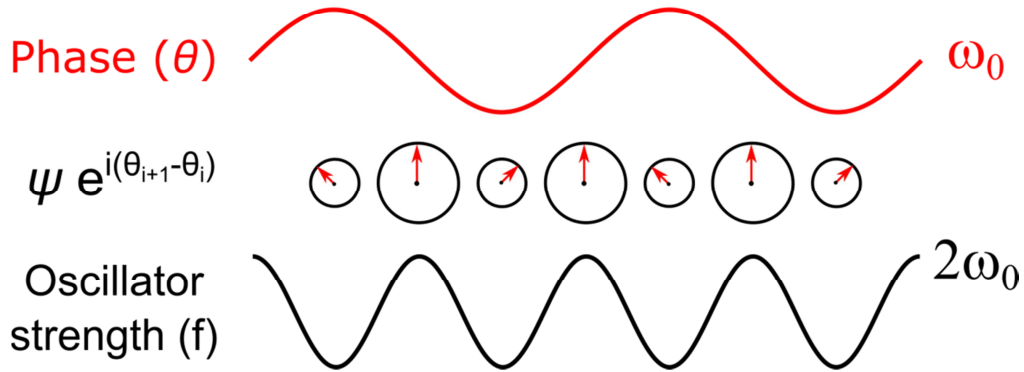

**Figure 1.** Schematic time-dependent representation of JPWs in linear and nonlinear regime, in presence of a driving field $E(t) = E_0 \sin(\omega_{JP0} t)$. Red arrows indicate the Josephson phase while the corresponding oscillator strength $f$ is represented by the black circle area. A JPW in linear regime consists of small amplitude modulations of $\theta_{i,i+1}$ at constant oscillator strength $f \sim \omega_{JP0}^2$. In nonlinear regime, the Josephson phase oscillates at $\omega_{JP0}$, whereas $f$ is modulated at $2\omega_{JP0}$.

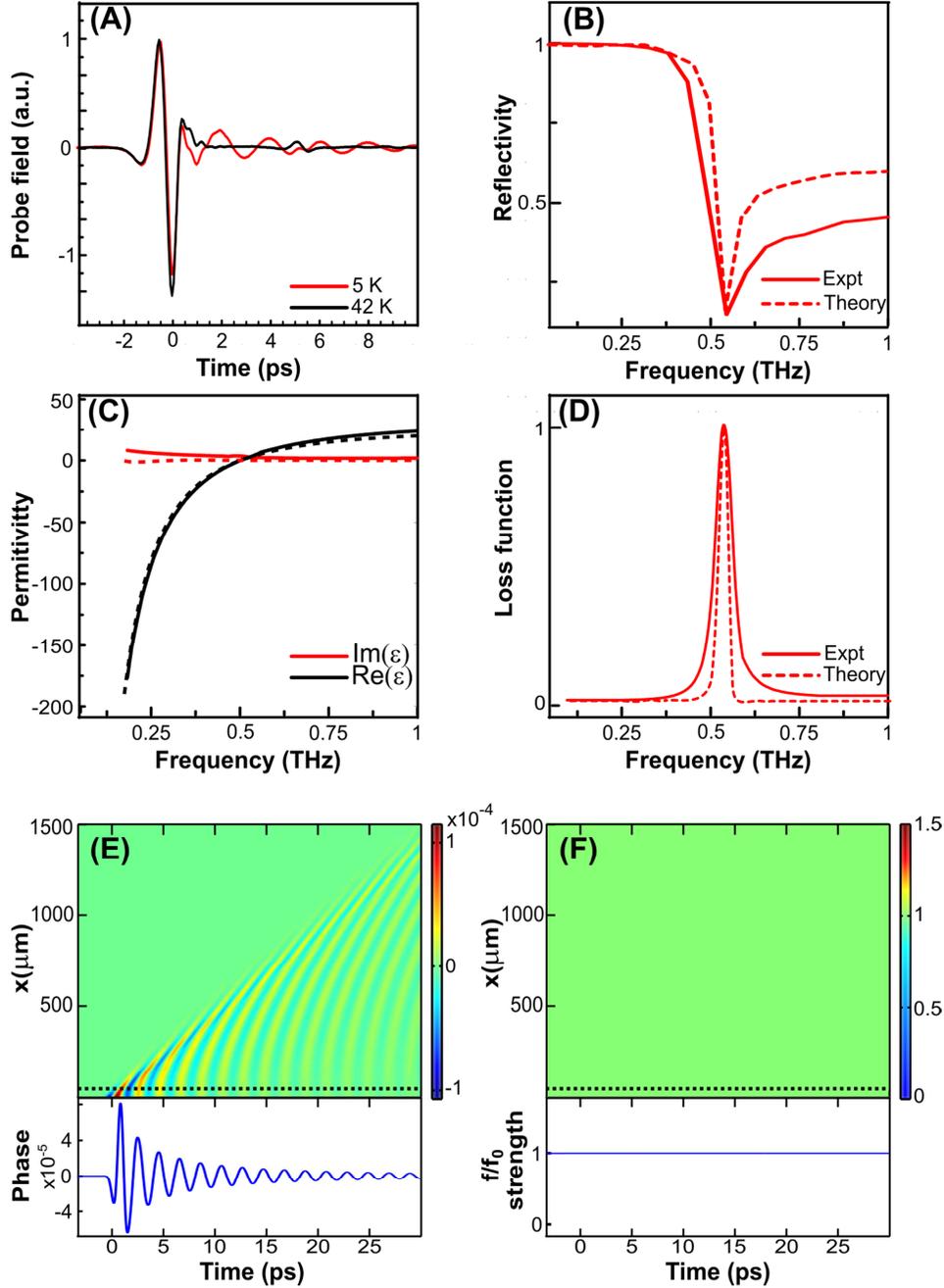

**Figure 2.** Linear JPWs in LBCO$_{9.5}$. (**A**) E($\tau$) measured in absence of pump field both above and below T$_c$ = 32 K. (**B**) Frequency-dependent, *c*-axis reflectivity at T = 5 K (solid line), extracted from the E($\tau$) trace of panel (**A**). (**C**) Corresponding real and imaginary part of the complex permittivity and (**D**) energy loss function (solid lines). Dashed lines in (**B-D**) were calculated by numerically solving the sine-Gordon equation in linear regime. (**E**) Phase $\theta_{i,i+1}(x,t)$ and (**F**) corresponding oscillator strength $f$ (no change) induced by a weak probe THz field, also determined from the sine-Gordon equation in linear regime. Horizontal dotted lines indicate the spatial coordinate $x$ at which the line cuts are displayed (lower panels).

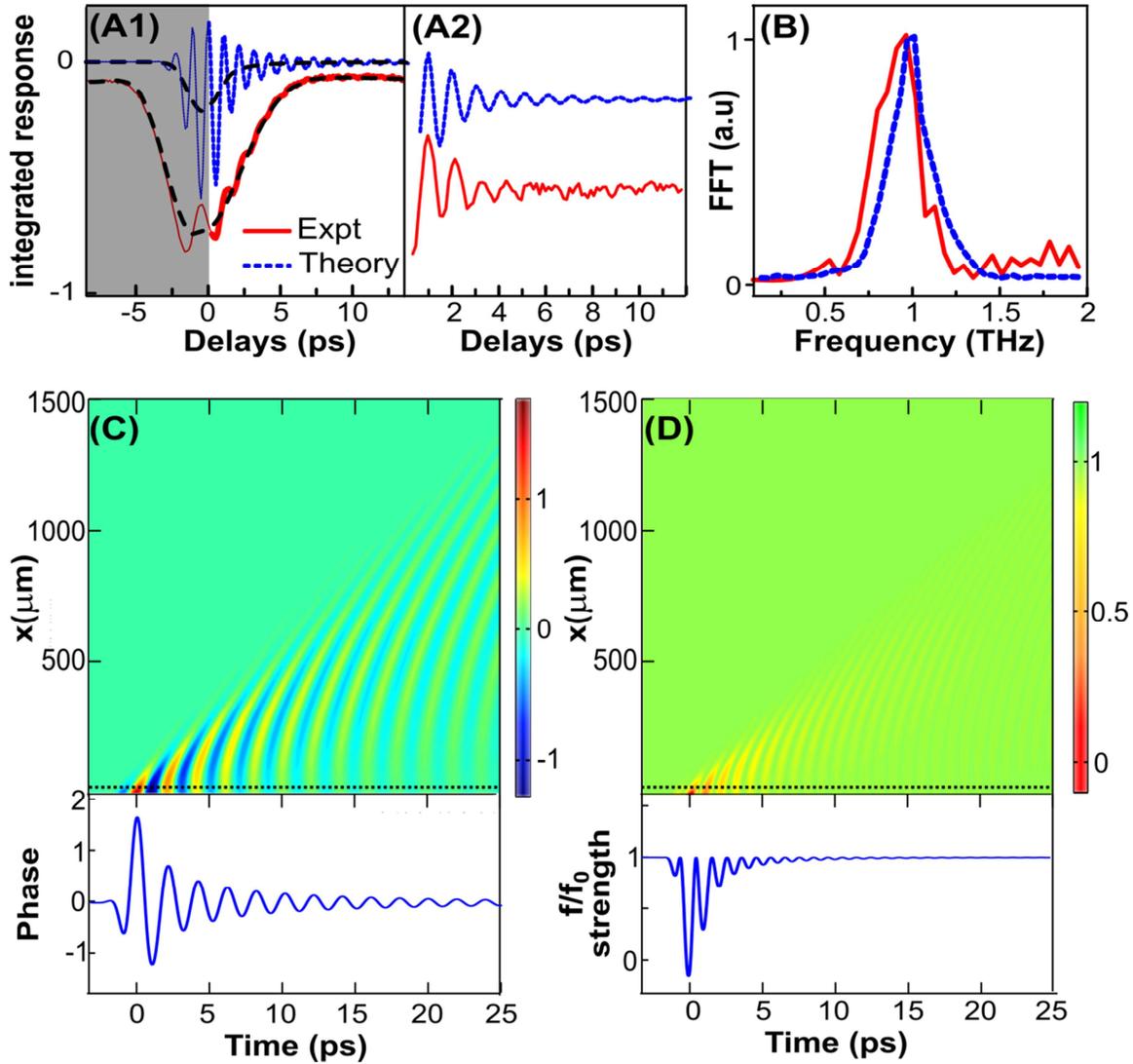

**Figure 3.** Nonlinear JPWs in $LBCO_{9.5}$. (**A1**) Spectrally integrated pump-probe response. Experimental data are displayed along with calculations based on the sine-Gordon equation in nonlinear regime. Dashed lines indicate the background which was subtracted to obtain the oscillatory component shown in (**A2**). (**B**) Fourier transform of the extracted oscillations, showing a peak at ~1 THz. (**C**) Phase $\theta_{i,i+1}(x,t)$ and (**D**) corresponding oscillator strength $f$ (normalized by the equilibrium value) induced by a strong THz pump field, as determined by numerically solving the sine-Gordon equation in nonlinear regime. Horizontal dotted lines indicate the spatial coordinate $x$ at which the line cuts are displayed (lower panels).

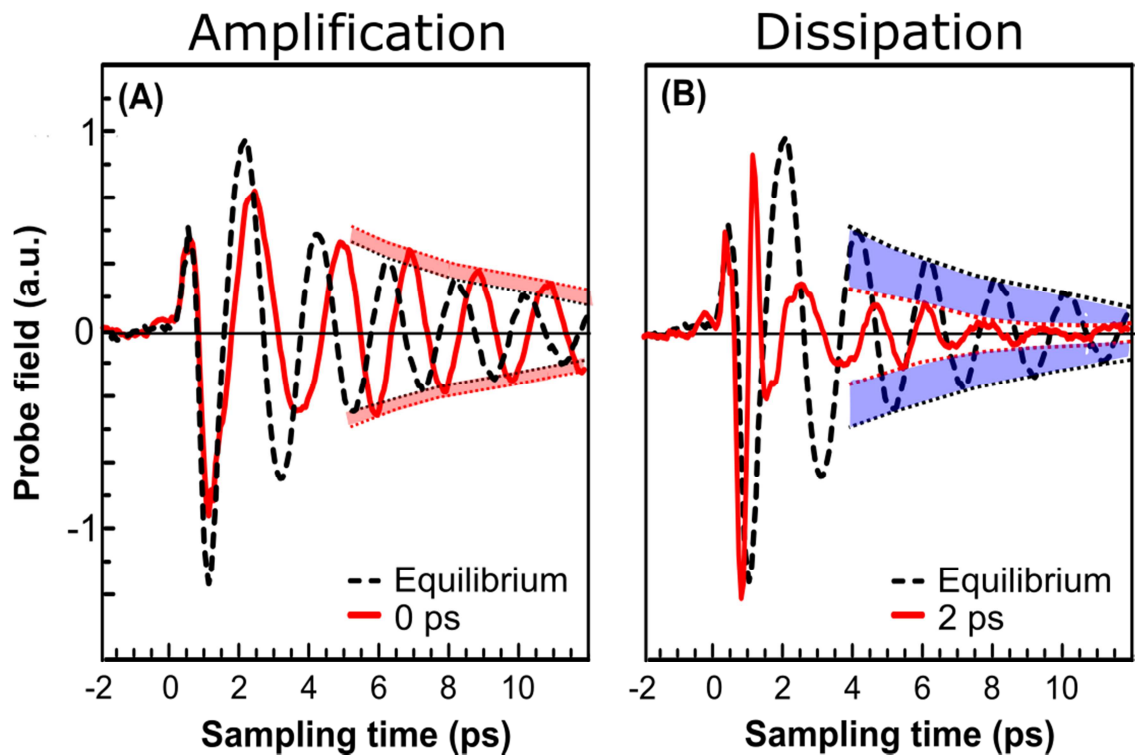

**Figure 4.** E(t,τ) traces measured by scanning the electro-optic sampling internal delay τ at selected pump-probe delays $t = 0$ ps and $t = 2$ ps. Data are shown along with the same quantity measured at equilibrium (pump off). Shaded regions in (**A**) and (**B**) indicate amplification and suppression of the JPW amplitude, respectively.

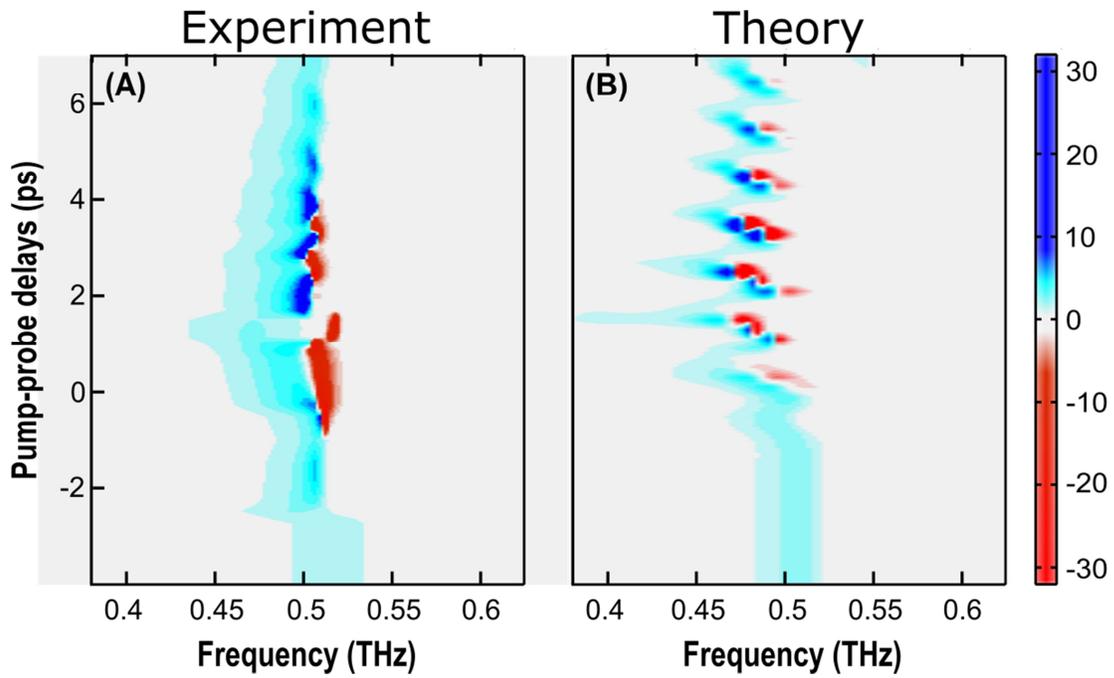

**Figure 5:** Time-delay and frequency dependent loss function $L(t,\omega)$ determined (**A**) experimentally and (**B**) by numerically solving the sine-Gordon equation in nonlinear regime. Note that the equilibrium $L(\omega)$ has been multiplied by a factor of 5.

# Parametric Amplification of
# a Terahertz Quantum Plasma Wave

S. Rajasekaran[1], E. Casandruc[1], Y. Laplace[1], D. Nicoletti[1], G. D. Gu[4], S. R. Clark[3,1], D. Jaksch[2,5], A. Cavalleri[1,2]

[1]Max Planck Institute for the Structure and Dynamics of Matter, Luruper Chaussee 149, 22761 Hamburg, Germany
[2]Department of Physics, Oxford University, Clarendon Laboratory, Parks Road, OX1 3PU Oxford, United Kingdom
[3]Department of Physics, University of Bath, Claverton Down, BA2 7AY, Bath United Kingdom
[4]Condensed Matter Physics and Materials Science Department, Brookhaven National Laboratory, Upton, New York, USA
[5]Centre for Quantum Technologies, National University of Singapore, 3 Science Drive 2, Singapore 117543, Singapore

# Supplementary Material

## S1. Josephson equation as Mathieu equation

A Josephson junction can be approximated with an LC circuit model. By equating the capacitive current $\left(I_c = C \frac{\partial V}{\partial t}\right)$ to the inductive tunneling current $\left(-I_L = -I_0 \sin \theta_{i,i+1}(t)\right)$ and using then the second Josephson equation $\left(\frac{\partial [\theta_{i,i+1}(t)]}{\partial t} = \frac{2eV}{\hbar}\right)$ we obtain the temporal dependence of the Josephson phase $(\theta_{i,i+1}(t))$ as

$$-\frac{\varepsilon_r}{c^2}\frac{\partial^2 \theta_{i,i+1}(x,t)}{\partial t^2} = \frac{\omega_p^2 \varepsilon_r}{c^2} \sin \theta_{i,i+1}(x,t) \tag{1}$$

where $\varepsilon_r$ is the dielectric permittivity of the Josephson junction, $c$ the speed of light, $e$ the electronic charge, $I_0$ the critical current, $C$ the capacitance of the junction, and

$$\omega_p^2 = \omega_0^2 = \frac{2 I_0 e}{\hbar C}.$$

The equation of motion of the Josephson phase with damping ($\gamma$) therefore reads

$$-\frac{1}{\gamma}\frac{\partial \theta_{i,i+1}(x,t)}{\partial t} - \frac{\varepsilon_r}{c^2}\frac{\partial^2 \theta_{i,i+1}(x,t)}{\partial t^2} = \frac{\omega_p^2 \varepsilon_r}{c^2} \sin \theta_{i,i+1}(x,t) \tag{2}$$

In a perturbed state in which the oscillator strength is modified as

$$f(t) \sim \omega_p^2(t) \approx \omega_0^2 \left(1 - \frac{\theta_0^2 + \theta_0^2 \cos(2\omega_0 t)}{4}\right) \tag{3}$$

the time dependence of the Josephson phase is described by

$$\frac{\varepsilon_r}{c^2}\frac{\partial^2 \theta_{probe}(x,t)}{\partial t^2} + \frac{1}{\gamma}\frac{\partial \theta_{probe}(x,t)}{\partial t} + \frac{\omega_0^2 \varepsilon_r}{c^2}\left(1 - \frac{\theta_0^2 + \theta_0^2 \cos(2\omega_0 t)}{4}\right)\theta_{probe}(x,t) = 0 \tag{4}$$

We note that Eq. (4) is a damped Mathieu equation of the form

$$\frac{\partial^2 \theta_{probe}(x,t)}{\partial t^2} + \beta \frac{\partial \theta_{probe}(x,t)}{\partial t} + (a - \alpha \cos(2\omega_0(t)))\theta_{probe}(x,t) = 0 \qquad (5)$$

where $a = \left(1 - \frac{\theta_0^2}{4}\right)\omega_0^2$, $\alpha = \frac{\theta_0^2 \cos(2\omega_0 t)}{4}\omega_0^2$ and $\beta = \frac{c^2}{\varepsilon_r \gamma}$.

## S2. Simulation of the nonlinear optical properties from the sine-Gordon equation

A Josephson junction with semi-infinite layers stacked along the $z$ direction (with translational invariance along the $y$ direction) can be modeled with the one-dimensional sine-Gordon equation[1,2]. Being $x$ the propagation direction, the Josephson phase evolution is described by:

$$\frac{\partial^2 \theta_{i,i+1}(x,t)}{\partial x^2} - \frac{1}{\gamma}\frac{\partial \theta_{i,i+1}(x,t)}{\partial t} - \frac{\varepsilon_r}{c^2}\frac{\partial^2 \theta_{i,i+1}(x,t)}{\partial t^2} = \frac{\omega_p^2 \varepsilon_r}{c^2} \sin\theta_{i,i+1}(x,t) \qquad (6)$$

The damping factor $\gamma$ is a fitting parameter used to reproduce the optical properties observed experimentally. In this section, we drop the subscripts for simplicity, i.e. we redefine $\theta_{i,i+1}(x,t) = \theta(x,t)$. The pump and probe THz fields impinge on the superconductor at the boundary $x = 0$. The Josephson phase evolution is therefore affected by the following boundary conditions at the vacuum-sample interface[3].

$$[E_i(t) + E_r(t)]_{x=-0} = E_c(x,t)|_{x=+0} = H_0 \frac{1}{\omega_{JPR}\sqrt{\varepsilon}} \frac{\partial \theta(x,t)}{\partial t}|_{x=+0}, \qquad (7)$$

$$[H_i(t) + H_r(t)]_{x=-0} = H_c(x,t)|_{x=+0} = -H_0 \lambda_J \frac{\partial \theta(x,t)}{\partial x}|_{x=+0}. \qquad (8)$$

The subscripts $i$, $r$, and $c$ denote the fields incident, reflected and propagating inside the cuprate, respectively. Here $H_0 = \Phi_0/2\pi D\lambda_J$, where $\Phi_0$ is the flux quantum $\left(\Phi_0 = \frac{hc}{2e}\right)$ and D is the distance between adjacent superconducting layers. The equilibrium Josephson Plasma Resonance (JPR) is an input parameter in the simulations, which is chosen to be that of $La_{1.905}Ba_{0.095}CuO_4$, i.e. $\omega_{JPR} = 0.5$ THz.

For fields in vacuum ($x < 0$), the Maxwell's equations imply

$$E_i - E_r = \frac{\omega\mu}{ck}(H_i + H_r) = H_i + H_r. \tag{9}$$

By combining Eq. (4) with Eq. (2) and (3) we obtain the boundary condition

$$\frac{2\sqrt{\varepsilon}}{H_0} E_i(t)|_{x=-0} = \frac{\partial\theta(x,t)}{\omega_{JPR}\partial t}\Big|_{x=+0} - \sqrt{\varepsilon}\frac{\partial\theta(x,t)}{\partial x/\lambda_J}\Big|_{x=+0}. \tag{10}$$

After solving the Josephson phase through Eq. (6) and Eq. (10), the reflected field is calculated from Eq. (7). The equilibrium reflectivity of the cuprate is obtained by computing the ratio between the Fourier transforms of the reflected field and a weak input field

$$r^{equilibrium}(\omega) = E_r^{equilibrium}(\omega)/E_i(\omega). \tag{11}$$

The complex optical properties are then calculated from $r^{equilibrium}(\omega)$. In particular, the equilibrium dielectric permittivity and loss function are computed as:

$$\varepsilon(\omega) = \left(\left(\frac{1-r^{equilibrium}(\omega)}{1+r^{equilibrium}(\omega)}\right)^2\right)$$

$$\text{Loss}(\omega) = -\text{Imag}\left(\left(\frac{r^{equilibrium}(\omega)+1}{r^{equilibrium}(\omega)-1}\right)^2\right)$$

For the pump-probe configuration, the input field is the sum of the pump and probe fields (with a defined delay between them):

$$E_i(t) = E_{pump}(t) + E_{probe}(t). \tag{12}$$

Correspondingly, the Josephson phase can be written as

$$\theta = \theta_{\text{pump}} + \theta_{\text{probe}}. \tag{13}$$

And the sine-Gordon equation (6) decomposes into two coupled equations

$$\frac{\partial^2 \theta_{pump}(x,t)}{\partial x^2} - \frac{1}{\gamma}\frac{\partial \theta_{pump}(x,t)}{\partial t} - \frac{\varepsilon_r}{c^2}\frac{\partial^2 \theta_{pump}(x,t)}{\partial t^2} = \frac{\omega_p^2 \varepsilon_r^2}{c^2}\sin\theta_{pump}(x,t)\cos\theta_{probe}(x,t) \tag{14}$$

$$\frac{\partial^2 \theta_{probe}(x,t)}{\partial x^2} - \frac{1}{\gamma}\frac{\partial \theta_{probe}(x,t)}{\partial t} - \frac{\varepsilon_r}{c^2}\frac{\partial^2 \theta_{probe}(x,t)}{\partial t^2} = \frac{\omega_p^2 \varepsilon_r^2}{c^2}\sin\theta_{probe}(x,t)\cos\theta_{pump}(x,t) \tag{15}$$

For a weak probe ($\theta \ll 1$), $\cos\theta_{\text{probe}} \approx 1$ and the effect of $\theta_{\text{probe}}$ on $\theta_{\text{pump}}$ can be neglected in Eq. (9). The phases $\theta_{\text{pump}}$ and $\theta_{\text{probe}}$ are calculated in two steps: (i) Eqs. (14) and (10) are solved with the driving field $E_i = E_{\text{pump}}$ to get $\theta_{\text{pump}}(x, t)$ and then (ii) Eq. (15) and (10) are solved by substituting $\theta_{\text{pump}}(x, t)$ with the input field $E_i = E_{\text{probe}}$, to obtain $\theta_{\text{probe}}(x, t)$ and the reflected probe field $E_r^{\text{perturb}}$. The perturbed reflectivity is given by

$$r^{\text{perturb}}(\omega, t) = E_r^{\text{perturb}}(\omega, t)/E_i(\omega). \tag{16}$$

The optical response functions of the perturbed material are extracted from the complex optical reflectivity $r^{\text{perturb}}$. For instance, the dielectric permittivity and loss function are calculated as:

$$\varepsilon(\omega) = \left(\left(\frac{1-r^{\text{perturb}}(\omega)}{1+r^{\text{perturb}}(\omega)}\right)^2\right)$$

$$\text{Loss}(\omega, t) = -\text{Imag}\left(\left(\frac{r^{\text{perturb}}(\omega,t)+1}{r^{\text{perturb}}(\omega,t)-1}\right)^2\right).$$

### S3. Pump spectrum

The electric field profile of the THz pump pulse (generated with the tilted pulse front technique in LiNbO$_3$) measured at the sample position is displayed in Fig. S1A alongside the corresponding frequency spectrum (Fig. S1B). This is peaked at ~0.5 THz, being therefore resonant with the JPR of LBCO$_{9.5}$ (see reflectivity edge in the blue curve of Fig. S1B). The input pump field used in the simulations is also displayed, both in time (Fig. S1A) and frequency domain (Fig. S1B).

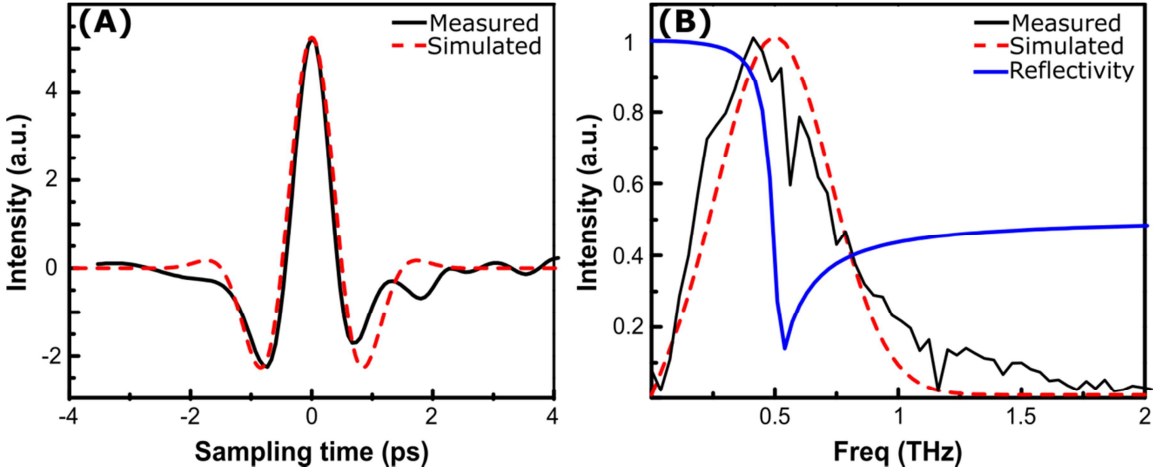

**Figure S1.** (**A**) Electro-optic sampling trace of the THz pump pulse measured at the sample position and (**B**) corresponding frequency spectrum. The *c*-axis equilibrium reflectivity of LBCO$_{9.5}$ at T = 5 K is also displayed. Dashed lines in both panels refer to the input pump field used in simulations. The ringing observed on the trailing edge of the pulse (black line in **A**) is due to narrow water absorption lines at ~0.5 THz and ~1.2 THz (see also corresponding spectrum in **B**). These can be ignored because, unlike the pump field scan shown in this figure, all other measurements reported in this paper have been performed under high vacuum condition (P = 10$^{-6}$ mbar).

## S4. Pump field dependence

The spectrally integrated "1D" pump-probe response[4,5] is displayed in Fig. S2 for different pump field strengths. A minimum field of ~30 kV/cm was required to induce a response of sufficient amplitude to be detected in our experiment.

The oscillatory behavior at twice the equilibrium JPR frequency was found to be only weakly dependent on the pump field strength. Note that pump-field-independent 2$\omega_{JP0}$ oscillations are only observed at $t > 0$, *i.e.* after the early-time dynamics ($t < 0$) dominated by perturbed free induction decay[6] (shaded region in Fig. S2).

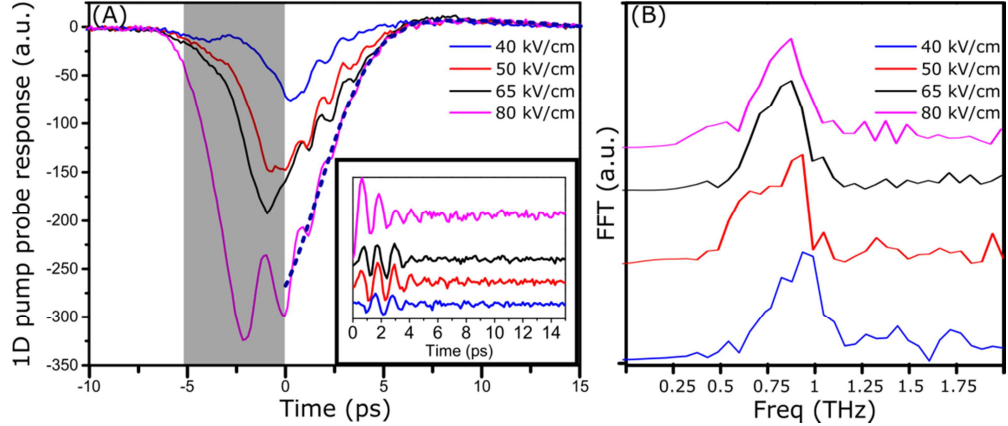

**Figure S2.** (**A**) Spectrally-integrated "1D" pump-probe response measured for different pump field strengths at a sample temperature T = 5 K. The dashed line is an example of background which was subtracted to extract the oscillatory components shown in the inset. The early time delay region, interested by perturbed free induction decay, is shaded in grey. (**B**) Normalized Fourier transforms of the oscillatory signals.

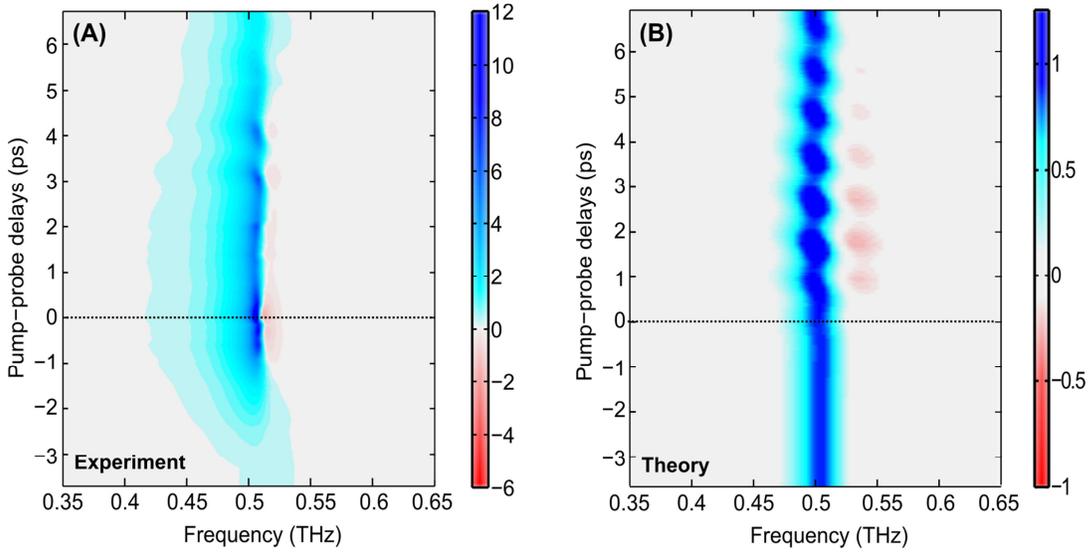

**Figure S3.** Time-delay and frequency dependent loss function determined (**A**) experimentally and (**B**) by numerically solving the sine-Gordon equation in nonlinear regime. The applied THz pump field is 40 kV/cm.

The time-delay and frequency dependent loss function measured with a pump field of 40 kV/cm is displayed in Fig. S3, along with the corresponding theoretical calculations. These can be compared with the data of Fig. 5 in the main text, which were taken with a higher pump field (~80 kV/cm). Remarkably, while the $2\omega_{JP0}$ oscillatory behavior is observed in both data sets, periodic amplification is only present with stronger pump field (consistently in both experiment and calculations).

This indicates that phase-sensitive amplification of Josephson Plasma Wave can be achieved only for THz pump field amplitudes above a threshold of ~70 kV/cm.

## S5. Temperature dependence

The equilibrium JPR shifts to lower frequencies with increasing temperature towards $T_c$. This is clearly shown in Fig. S4A, where the measured equilibrium reflectivity of LBCO$_{9.5}$ is reported at two different temperatures. The JPR exhibits a red shift from ~0.5 THz to ~0.35 THz upon increasing the sample temperature from 5 K to 30 K.

The temperature dependence of the spectrally integrated pump-probe response has also been determined experimentally (see Fig. S4B). As expected, the measured oscillations slow down with increasing $T$. Indeed their frequency reduces from ~ 1 THz at 5 K to ~0.75 THz at 30 K, scaling as $2\omega_{JP0}$.

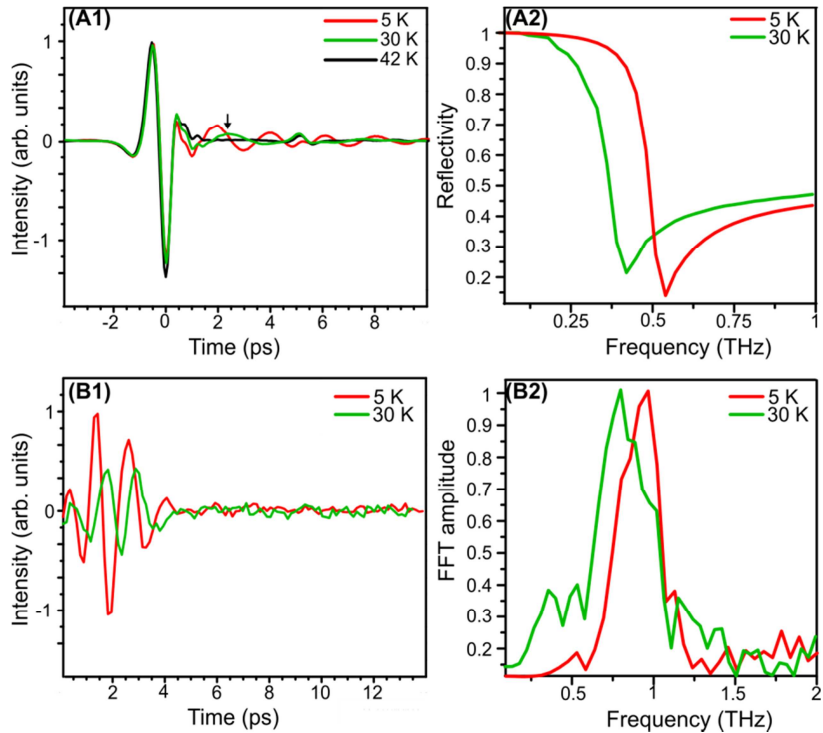

**Figure S4.** (**A1**) $E(\tau)$ measured in absence of pump field at different temperatures above and below $T_c$. (**A2**) Frequency-dependent reflectivity at T = 5 K and T = 30 K, extracted from the $E(\tau)$ trace of panel (**A1**). (**B1**) Oscillatory component of the spectrally-integrated "1D" pump-probe response, measured at T = 5 K and T = 30 K. (**B2**) Corresponding Fourier transforms of the oscillatory signals.